\documentclass[iop,apj,tighten,a4paper]{emulateapj}

\usepackage{amsmath,amstext,amssymb}
\usepackage[breaklinks,colorlinks,citecolor=blue,linkcolor=magenta]{hyperref}

\slugcomment{Accepted for publication in ApJ}
\shorttitle{SMBH Coalescence in Cosmological Galaxy Mergers}
\shortauthors{Khan et al.}

\begin{document} 

\title{Swift coalescence of supermassive black holes in cosmological mergers\\ of massive galaxies}

\author{Fazeel Mahmood Khan\altaffilmark{1}, Davide Fiacconi\altaffilmark{2}, Lucio Mayer\altaffilmark{2}, Peter Berczik\altaffilmark{3,4,5}, and Andreas Just\altaffilmark{5}}
\email{khanfazeel.ist@gmail.com}
\altaffiltext{1}{Department of Space Science, Institute of Space Technology, PO Box 2750 Islamabad, Pakistan}
\altaffiltext{2}{Center for Theoretical Astrophysics and Cosmology, Institute for Computational Science, University of Zurich, Winterthurerstrasse 190, CH-8057 Z\"{u}rich, Switzerland}
\altaffiltext{3}{National Astronomical Observatories and Key Laboratory of Computational Astrophysics, Chinese Academy of Sciences, 20A Datun Rd., Chaoyang District, 100012, Beijing, China}
\altaffiltext{4}{Main Astronomical Observatory, National Academy of Sciences of Ukraine, 27 Akademika Zabolotnoho St., 03680, Kyiv, Ukraine}
\altaffiltext{5}{Astronomisches Rechen-Institut, Zentrum f\"{u}r Astronomie der Universit\"{a}t Heidelberg, M\"{o}nchhofstrasse 12-14, 69120, Heidelberg, Germany}


\begin{abstract}
Supermassive black holes (SMBHs) are ubiquitous in galaxies with a sizable mass.
It is expected that a pair of SMBHs originally in the nuclei of two merging galaxies 
would form a binary and eventually coalesce via a burst of gravitational waves.
So far theoretical models and simulations have been unable to predict directly the SMBH 
merger timescale from ab-initio galaxy formation theory, focusing only on limited phases 
of the orbital decay of SMBHs under idealized conditions of the galaxy hosts. The predicted 
SMBH merger timescales are long, of order Gyrs, which could be problematic for future 
gravitational wave searches. Here we present the first multi-scale $\Lambda$CDM cosmological 
simulation that follows the orbital decay of a pair of SMBHs in a merger of two typical 
massive galaxies at $z\sim3$, all the way to the final coalescence driven by gravitational 
wave emission. The two SMBHs, with masses $\sim10^{8}$~M$_{\odot}$, settle 
quickly in the nucleus of the merger remnant. The remnant is triaxial and extremely dense 
due to the dissipative nature of the merger and the intrinsic compactness of galaxies at high 
redshift. Such properties naturally allow a very efficient hardening of the SMBH binary. The 
SMBH merger occurs in only $\sim10$~Myr after the galactic cores have merged, which is two 
orders of magnitude smaller than the Hubble time.
\end{abstract}

\keywords{black hole physics -- galaxies: interactions -- galaxies: kinematics and dynamics -- galaxies: nuclei -- gravitational waves -- methods: numerical}


\section{Introduction}\label{sec-intro}

Dual active galactic nuclei (AGNs) at kiloparsec to hundred parsec separations have been
detected \citep{comerford+13}, but at smaller separations there are only unconfirmed
candidates \citep{eracleous+12,graham+15}.
At the same time, the orbital decay of two supermassive black holes (SMBHs) at the center 
of merging galaxies \citep{begelman+80} has been theoretically studied with a variety of 
computer simulations, but always neglecting one or more important processes. 
Substantial work has focused on the gravitational interaction between the SMBHs 
and the stellar background, as it would be appropriate for a gas-free galaxy \citep{makino+04,vasiliev+15},
as well as on the dynamics of two SMBHs within a dissipative gaseous background, though
neglecting gravitational encounters with individual stars \citep{dotti+07,mayer+07,chapon+13}.
Furthermore, individual simulations, due to computational limitations, typically follow only a limited phase 
of the SMBH pair evolution, either before or after a Keplerian binary 
forms \citep{mayer+13}.
It has been noticed that SMBHs in stellar systems may stall at parsec separations as the loss cone 
is depleted, rendering the transfer of energy between the binary and the stellar background
inefficient \citep{makino+04,berczik+05}.
This has been dubbed the ``last parsec problem'' \citep{milosavljevic+01}.
However, it has been shown that the loss cone can be refilled if the potential of the galaxy 
has substantial deviations from sphericity \citep{khan+11,preto+11,vasiliev+15}.
Yet, extrapolating the hardening rates seen in these recent simulations to the gravitational wave 
(GW) dominated phase using analytical models, the resulting SMBH merger timescales are 
$\sim 1$~Gyr \citep{khan+12a,khan+12b}. 
The galaxy merger timescale is also of order a Gyr \citep{stewart+09}, suggesting that
the overall process takes a significant fraction of the age of the Universe. 
At $z>2$, when the lookback time is also of order a few Gyr, this would imply low SMBH coalescence 
rates, a potential problem for future GW experiments such as the Evolved Laser Interferometer 
Space Antenna (eLISA) \citep{amaro+13}.

If gas is present, such as in circumnuclear disks forming as a result of gas-rich mergers, the orbital
decay of the SMBHs proceeds on a faster track, leading to a hard binary with pc-scale separation in 
$\sim1-100$~Myr, depending on the  clumpiness of the interstellar medium 
\citep{mayer+07,chapon+13,fiacconi+13,roskar+15}.
However, the drag may become inefficient at smaller separations, 
potentially resulting in a stalling binary \citep{chapon+13,mayer+13}.

The results of all these simulations depend strongly on the mass distribution and properties of the stars 
and interstellar gas of the circumnuclear region, which are inherited from idealized initial conditions.
Therefore it is still unclear how and at what pace the orbital decay of SMBHs proceeds in realistic galaxy mergers.
In order to make progress, in this paper we have carried out the first ab-initio calculation of a SMBH merger 
that starts from a galaxy merger pinpointed in a state-of-the-art cosmological hydrodynamical simulation. 
We follow the SMBHs all the way to the final spiral-in phase with the aid of post-Newtonian corrections \citep{blanchet+06}.


\section{Galaxy Merger Simulation} \label{mer-sim}

We identify the merger of two massive galaxies at $z \sim 3.5$ in the Argo cosmological hydrodynamical simulation \citep{feldmann+15,fiacconi+15}.
The simulation follows the formation of a group-sized halo with mass $\approx2\times10^{13}$~M$_{\odot}$ at $z=0$, and includes gas cooling, star formation (SF) and a supernovae feedback models that have been shown to produce realistic galaxies at a variety of mass scales \citep{governato+10,guedes+11}.
The halo evolves in a mildly over-dense region and its virial mass is close to the characteristic scale $M^{\star}$ of the halo mass function at low $z$, suggesting that it should be a common host for massive galaxies \citep{feldmann+15,fiacconi+15}. 
In lower resolution simulations the group hosts a central galaxy with properties typical of massive early-types at $z=0$ \citep{feldmann+10}.

\begin{figure*}
\epsscale{1.1}
\plotone{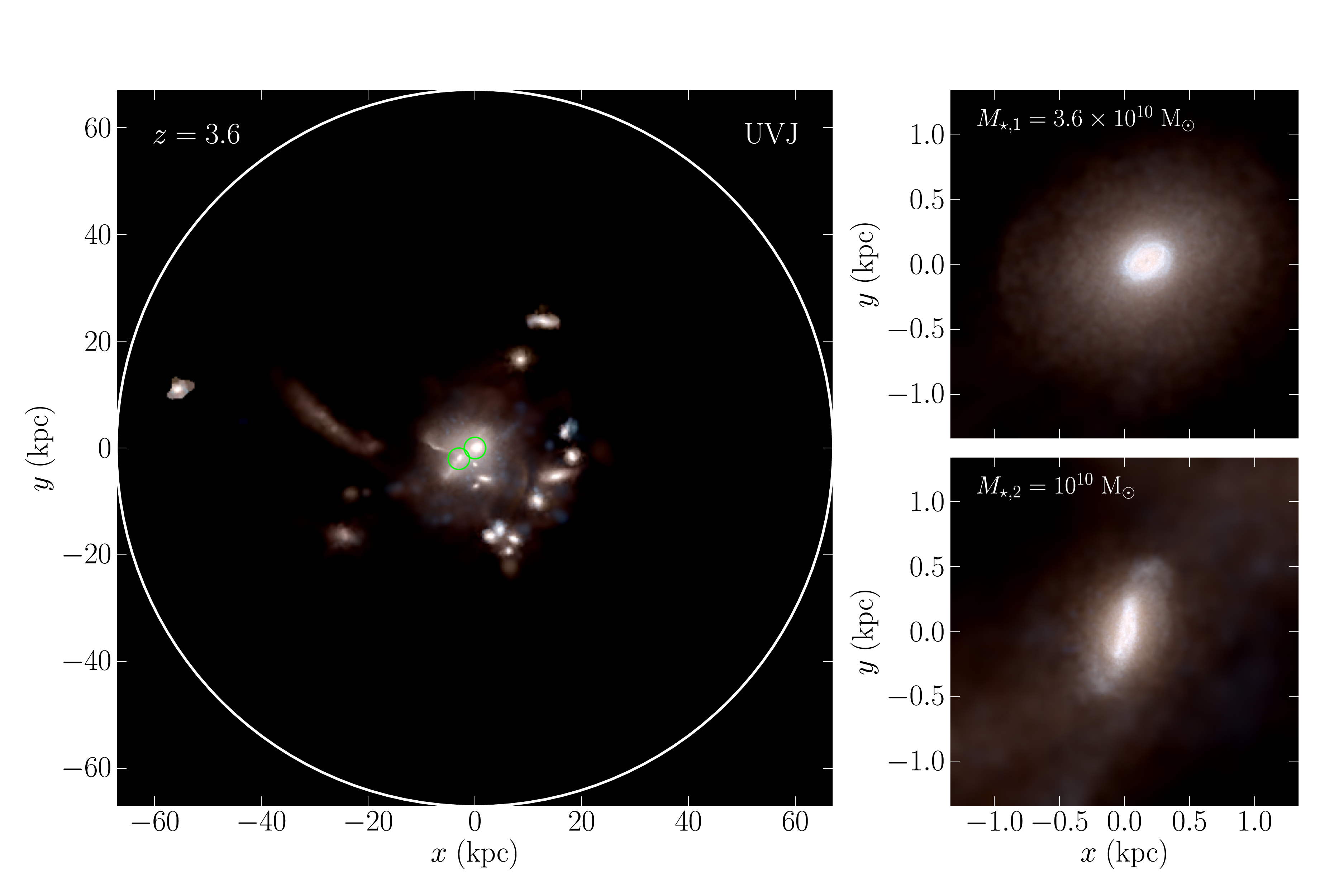}
\caption{Group environment of the galaxy merger.
The left panel shows a mock UVJ map of the galaxy group at $z=3.6$.
The white circle marks the virial radius of the group halo, while the green circles 
mark the merging galaxies.
The upper-right and lower-right panels show a zoom-in on the central galaxy of the 
group and the interacting companion, respectively.
Lengths are in physical coordinates.
\label{fig1}}
\end{figure*}

The central galaxy of the Argo simulation undergoes its last major merger (with a stellar mass ratio $\sim 0.3$) at $z \simeq 3.5$.
The merger involves two disk-like galaxies in a nearly parabolic (slightly hyperbolic) orbit, with their stellar spins misaligned by $\approx 67^{\circ}$.
Such a configuration is typical for major mergers in $\Lambda$CDM cosmology \citep{khochfar+06}.
The two galaxies have stellar masses $M_{\star,1} \approx 3.6 \times 10^{10}$~M$_{\odot}$ and $M_{\star,2} \approx 10^{10}$~M$_{\odot}$, and gas fractions $f_1 \approx 7.7\%$ and $f_2 = 11.5\%$, respectively.
The two galaxies and their group environment are shown in Figure \ref{fig1} when they are about to merge.

The cosmological simulation does not originally contain any SMBH, and its 
resolution would not allow us to probe the evolution of a BH binary.
Therefore, we increase the resolution by performing 
static particle splitting \citep{mayer+07,roskar+15}.
Specifically, we extract from the cosmological simulation a spherical region with radius $\sim13.5$~kpc 
at $z=3.6$ that encompasses the two galaxies and part of their environment.
We check that the average dynamical time of such region is $\gtrsim 100$~Myr, which is larger than the 
dynamical time in the central region and the simulation time that we target, about a few tens of Myr.
At this stage, the cores of the two galaxies are at a separation of $\lesssim 4$~kpc.
We then split all particle species in 8 child particles with masses 8 times smaller and the same 
velocity of the parent particle, thus conserving mass and linear momentum exactly, and angular 
momentum at the kernel level.
Thermodynamic properties of gas particles (i.e. density and temperature) are interpolated among the 
child particles \citep{roskar+15}, while child stellar particles maintain the properties of their 
parents (e.g. the age).
After the splitting, the simulation contains 9~452~581 stellar particles, 1~088~920 gas particles, 
and 1~669~922 dark matter particles with masses $6.4\times10^3$~M$_{\odot}$, $2\times10^4$~M$_{\odot}$, 
and $10^5$~M$_{\odot}$, respectively.
We reduce to $\epsilon=5$~pc the gravitational softenings\footnote{The gravitational softenings of the Argo simulation are $120$~pc and $250$~pc for baryons and dark matter, respectively.} of gaseous and stellar particles to increase 
the spatial resolution, while the dark matter softening is reduced by a factor $8^{1/3}=2$ to maintain 
the local density and the smooth gravitational field.
We extensively tested this procedure by running twin simulations with $\epsilon=15$ and 50~pc; we checked 
that (i) no spurious effects on scales larger than the original softening were introduced, and (ii) the 
dynamics of the introduced SMBHs (see below) converged down to the adopted softenings
 (see Section \ref{sec_appendix} for quantitative details).

During the splitting procedure, we introduce two SMBHs at the local minima of the gravitational 
potential of the galactic cores.
We assign to the SMBHs the mass-weighted average velocity of all the particles within 250~pc from 
their positions.
We checked that the velocities do not depend strongly on the size of the regions that we choose.
Since the two galaxies are relatively gas-poor at $z\sim3.5$ and we aim to continue the simulation 
for $\sim20-40$~Myr, we neglect mass accretion and feedback from the SMBHs, that are thus treated 
as collisionless particles.
The SMBHs have the same softening as stellar and gas particles.
We finally choose their masses according to local scaling relations \citep{ferrarese+00,tremaine+02,mcconnell+13,sco13,kor14}.
We used \citep{mcconnell+13} to determine the SMBH masses after measuring the average velocity dispersion 
within one half-mass radius for each galaxy before performing particle splitting.
The resulting masses are $M_{\bullet,1}=3\times10^{8}$~M$_{\odot}$ and 
$M_{\bullet,2}=8\times 10^{7}$~M$_{\odot}$, with a mass ratio $M_{\bullet,1}/M_{\bullet,2}=3.75$.
Though using the local scaling relations for the SMBH masses is formally inappropriate for high-$z$ 
galaxies, this might result to be a conservative choice because there are both observational 
\citep{merloni+10,trakhtenbrot+15} and theoretical \citep{degraf+15} hints that the normalization of 
those relations increases at high-$z$, i.e. galaxies host 
proportionally larger SMBHs.

\begin{figure*}
\epsscale{1.1}
\plotone{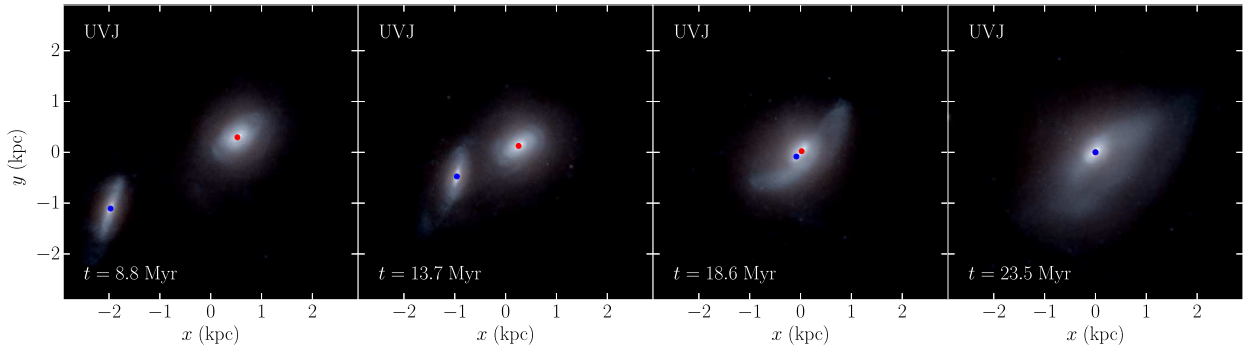}
\caption{From left to right: time evolution of the galaxy merger after the beginning of the re-sampled, 
higher-resolution simulation.
Each panel shows a mock UVJ photometric image of the merger, and the red and blue dots mark the position 
of the primary and secondary BH, respectively.
Lengths are in physical coordinates.
\label{fig2}}
\end{figure*}

After setting-up the initial conditions as described, we simulate the final stages of the galaxy merger 
until the separation of the two SMBHs is about the resolution. 
We use the {\sc gasoline} code \citep{wadsley+04}, but including additional sub-resolution physics.
Specifically, we add the gas radiative cooling from metal lines, a pressure floor to avoid spurious 
fragmentation, and an equilibrium temperature-density relation for gas denser than 0.1~H~cm$^{-3}$ to 
model the optically-thick phase, calibrated on 2D radiative transfer simulations in typical starburst 
conditions \citep{spaans+00,roskar+15}.
We also increase the density threshold to form star to 1000~H~cm$^{-3}$ and we reduce the temperature 
threshold to 300~K, to account for the new cooling.

Figure \ref{fig2} shows different stages of the merger of the two galaxies in our simulation after we 
perform the particle splitting.
The Figure reveals that the two galaxies are flattened and disk-like.
The remnant has an elongated shape out to a few kpc soon after the merger.

\begin{figure*}
\epsscale{1.1}
\plotone{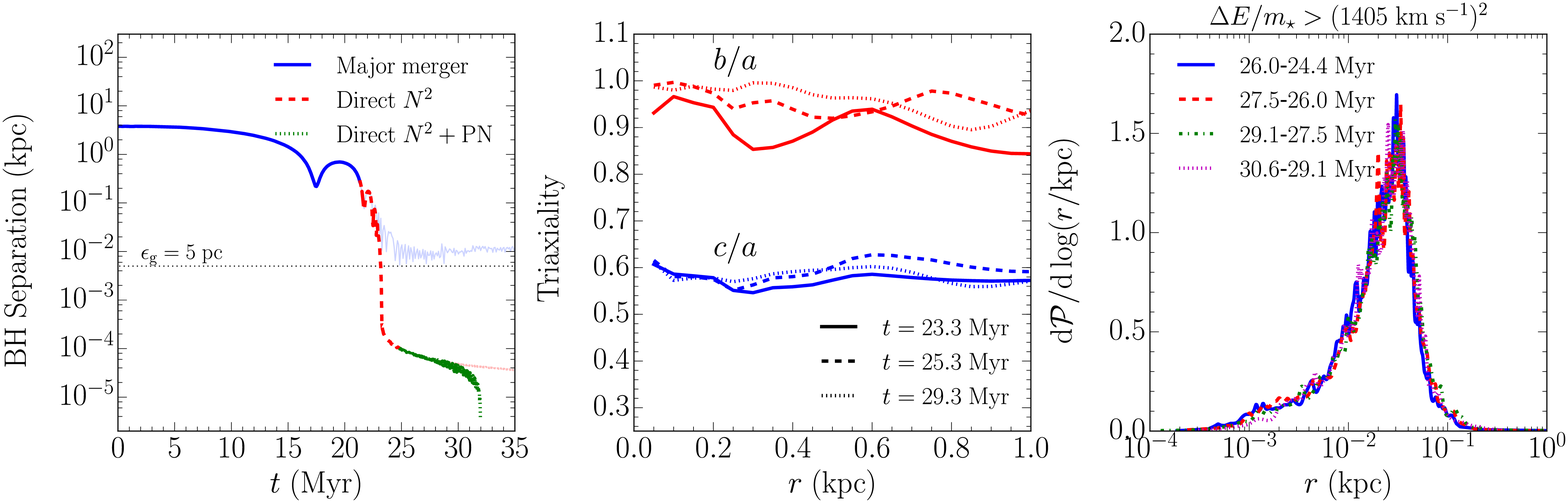}
\caption{Left panel: time evolution of the separation between the SMBHs.
Blue-solid, red-dashed, and green-dotted lines show the evolution during the hydrodynamical, re-sampled simulation of the merger, the direct $N$-body calculation, and after having introduced post-Newtonian corrections, respectively.
Thin and light versions of the same lines refer to the continuation of the respective simulations.
The horizontal dotted line marks the gravitational softening of the hydrodynamical simulation.
Central panel: radial profiles of the ratio $b/a$ (red) and $c/a$ (blue) between the principal axes of the moment of inertia tensor ($c \leq b \leq a$) at different times: 23.3~Myr (solid), 25.3~Myr (dashed), and 29.3~Myr (dotted).
Right panel: probability density function of the radial distance from the center of the merger remnant for the stellar particles that have interacted with the central binary across 26-24.4~Myr (blue, solid), 27.5-26~Myr (red, dashed), 29.1-27.5~Myr (green, dot-dashed), and 30.6-29.1~Myr (magenta, dotted).
\label{fig3}}
\end{figure*}


\section{Supermassive Black Hole Binary Evolution} \label{BBH-evo}

The left panel of Figure \ref{fig3} describes the orbital evolution of the two SMBHs starting form 
$\sim 4$~kpc till the final coalescence.
As the two galaxies merge, their SMBHs sink in the remnant surrounded by stellar cusps bound to them.
The orbital decay is governed by dynamical friction of the stellar cusps against the stellar, gas 
and dark matter background originating from the merger of the two hosts.

\begin{figure}
\epsscale{1.1}
\plotone{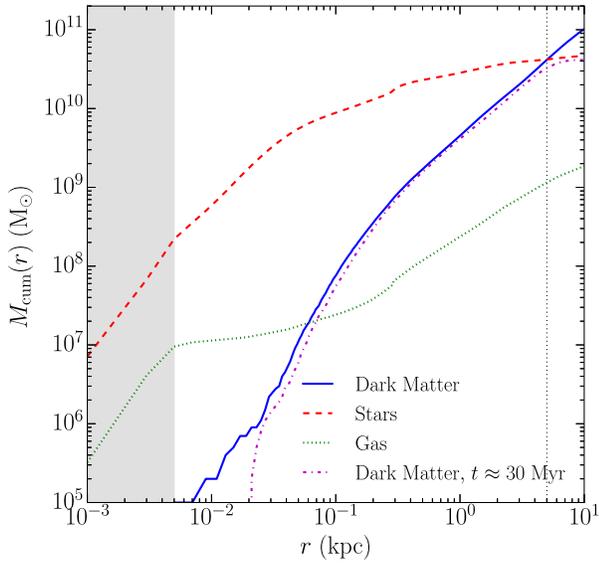}
\caption{Enclosed mass profile of dark matter (blue, continuous), stars (red, dashed) and gas (green, dotted) at $t\sim 21.5$~Myr 
when we select the inner 5~kpc (vertical dotted line) for the direct $N$-body simulation.
The gray area marks $\epsilon=5$~pc.
The magenta dot-dashed line shows the dark matter profile at $t\approx30$~Myr, which is not modified by our truncation.
\label{fig4}}
\end{figure}

During the final stage of the merger (i.e. at $t\approx20$~Myr after the particle splitting) the 
merger remnant is gas poor (gas fraction $\sim5\%$) owing to gas consumption by SF.
Stars dominate the enclosed mass out to $\sim3$~kpc and provide the dominant contribution to the 
dynamical friction exerted by the background.
Figure \ref{fig4} shows the mass distribution of the individual components when 
the separation of the two SMBHs reaches about 300 pc, i.e. $\approx 21.5$~Myr after the particle 
splitting.
The stellar mass is almost 2 orders of magnitude larger than the gas one over all spatial scales 
except in the central 10~pc, where the difference is about a factor of 20.

Then, we extract a spherical region of $5$~kpc at $t\sim 21.5$~Myr after particle splitting 
around the more massive SMBH to initialize a direct $N$-body simulation containing in total\
$\sim6\times10^{6}$ particles.
We treat the remaining gas particles in the selected volume as stars, since they are sub-dominant
in mass.
Almost the entire stellar mass is within 5~kpc, so an artificial cut-off at 5~kpc shall not introduce 
significant changes in stellar mass profile in the inner region for follow up evolution.
However, at truncation separation, the dark matter has a steeply rising mass profile; we compare it
with a later snapshot during the $N$-body evolution at $t\approx 30$~Myr.
We do not observe a noticeable evolution from outer to inner region.

We further evolve the selected region using the high-performance $\phi$-GPU code~\citep{berczik+11}.
At the end of our previous galaxy merger simulation, stellar and gas particles have a softening of 5 pc 
and dark matter particles have a softening of 150 pc.
We start our direct $N$-body run by decreasing the stellar softening to 0.1 pc while keeping the dark matter 
particles softening unchanged to avoid two-body relaxation effects as the latter have a relatively large mass.
Figure \ref{fig4} shows that the mass of the dark matter component does not increase in the 
central region during the whole evolution period.
The softening parameter for the force calculation between the two black holes is set to 0.
In order to calculate the softening between different particle species we employ the following criterion:
\begin{equation} \label{eq_soft}
\epsilon_{ij} ^2 = (\epsilon_{i}^{2} + \epsilon_{j}^{2})/2,
\end{equation}
where $\epsilon_{\bullet} = 0$ for both black holes, $\epsilon_{\star} = 0.1$ pc for stars and $\epsilon_{\rm dm} = 125$ pc 
for dark matter particles.
In star-black hole interactions we further reduce the softening to  0.007 pc, which is smaller than the semi-major axis of 
the binary when the gravitational wave emission dominates (Figure \ref{fig3}, left panel).
In order to take into account energy loss by gravitational wave emission, we incorporate post-newtonian terms up to 3.5 in the
equation of motion of the binary SMBHs \citep{blanchet+06}.

Dynamical friction efficiently shrinks the separation between the two SMBHs and they form a binary 
once individual cusps merge at $t \sim 23.5$~Myr.
The separation drops rapidly to $\sim 0.3$~pc in less than 1 Myr, owing to the high nuclear density, 
until the binary gets hard and dynamical friction becomes inefficient.
The subsequent phase of the decay is dominated by three-body encounters between the binary and the 
surrounding stars.
This phase is the longest, taking $\approx8$~Myr before the separation decreases to $\sim0.01$~pc, 
at which point GW emission takes over and brings the SMBHs to rapid 
coalescence in 2~Myr 
(Figure \ref{fig3}).
Figure \ref{fig3} also shows that post-Newtonian terms are crucial for the sinking of the binary 
already at a separation of $\gtrsim0.03$~pc.
The orbital decay rate in the post-Newtonian phase is in rough agreement with simple semi-analytical 
predictions based on orbit-averaged expressions \citep{peters+63}, which do not take into account the
contributions from higher order terms.

We define the coalescence time, and stop the simulations, when the separation is  $<4 (r_{\rm s,1} + r_{\rm s,2})$,
where $r_{{\rm s},j}$ is the Schwarzschild radius of the $j$-th BH, as 
following the evolution further 
would require a fully relativistic treatment.
The coalescence, counting from the merger of the two cusps at $t \sim 23.5$~Myr, takes less than 10~Myr, 
which is roughly two orders of magnitude faster than previous decay time estimates inferred for non-cosmological,
purely stellar hosts (when rescaled to nearby galaxies) \citep{khan+12b,khan+13}.
The preceding large-scale approach and merger of the two galaxies lasts $\sim 200$~Myr, so that the 
overall process is completed in significantly less than the lookback time at $z \sim 3.5$.

The shape of the merger remnant over time is shown in the central panel of Figure \ref{fig3}.
It was obtained by measuring the moments of inertia tensor of a homogeneous ellipsoid.
The remnant is clearly triaxial at all times.
The right panel of Figure \ref{fig3} shows the distribution of the mean radial distances 
of the stars that contribute to the change in the binding energy of the binary at
different times.
Those have been identified statistically as the stars that undergo large specific total energy change between two 
subsequent snapshots, 
$\Delta E / m_{\star} > (1405~{\rm km~s^{-1}})^2$.
This absolute threshold depends on the binary properties only, 
$\Delta E / m_{\star} \approx G~\mu_{\bullet}~C~a^{-1}$,
where $\mu_{\bullet}$ is the binary reduced mass, $a$ is the binary separation, and 
$C\approx2$, as inferred from three-body encounter simulations \citep{hills+83,quinlan+96}.
This criterion allows us to select only the stars involved in encounters with the SMBH binary, 
because the energy changes due to the large scale evolution of the system and to two-body 
encounters with other stars and dark matter particles are small compared to the adopted 
threshold.
Most of those stars come from 10-100~pc, which is at quite far from the BH binary 
sitting at the center of the remnant. 
This shows that the loss cone is efficiently refilled from stars on plunging orbits.

Finally, we stress that, at the resolution that we are employing, our results on the binary
evolution during the phase dominated by stellar encounters are robust.
Indeed, previous tests in flat, rotating, and triaxial systems have shown that the hardening 
rate is almost independent of the number of particles when above $\sim10^6$
\citep{berczik+06,khan+11,preto+11,khan+13,holley+15}.
Since the star particles in our run are $\sim5.5\times10^{6}$, we expect that we reach a regime 
of statistical convergence where fewer encounters with more massive particles produce a total 
energy exchange comparable to that occurring with many more encounters with lighter particles.

\begin{figure}
\epsscale{1.1}
\plotone{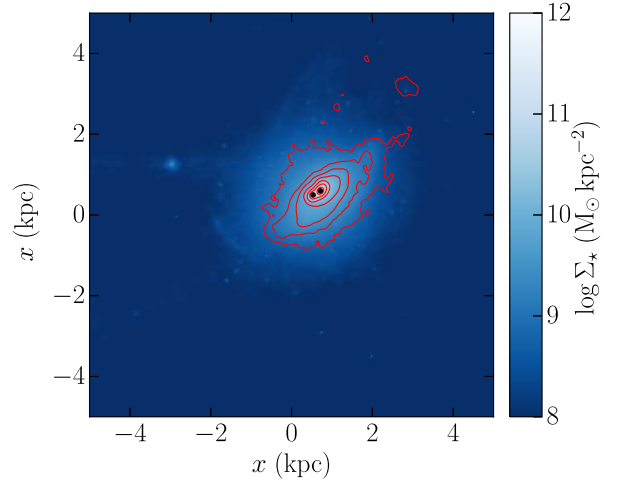}
\caption{Surface density map of the stellar component at the time of the beginning of the
$N$-body simulation. Red continuous lines represent isocontours of stars younger
than $\sim 22.5$~Myr (i.e. formed from the beginning of the resampled merger
simulation) from $5 \times 10^{8}$ to $5 \times 10^{11}$~M$_{\odot}$~kpc$^{-2}$
with steps of 0.5 dex. The black dots denote the positions of the two black holes.
\label{fig5}}
\end{figure}


\section{Discussion and Conclusions}

In this paper, we discuss the first multi-scale simulation that probes the evolution of a SMBH binary
forming within a cosmological major merger, all the way down to the coalescence driven by the emission of GWs.
We start from the Argo cosmological simulation, where we identify and re-simulate at higher resolution a major
merger between two massive galaxies at $z\approx3.5$.
Gas dissipation before the merger is instrumental in creating the conditions for the rapid orbital decay of the SMBHs, 
which is our key finding.
Indeed, the high central stellar density in the remnant is the result of gas inflows in the inner $\lesssim500$~pc 
due to prior cosmological gas accretion and  mergers \citep{feldmann+10}.

These effects cannot be accounted for in idealized galaxy mergers, rather require cosmological simulations as those 
employed here. 
The dense remnant not only causes strong dynamical friction by both gas and stars in the early decay phase, but provides 
an abundant reservoir of stars in the nuclear region that interact with the binary in the late phases of the decay. 

At the same time, triaxiality is necessary to avoid the loss cone problem and is a natural result of mergers between 
non-spherical galaxy progenitors.
Since massive galaxies comprise a large fraction of star forming galaxies at high $z$, many with massive disk-like 
components \citep{wisnioski+15}, the pre-merger conditions in our simulation should be typical of the progenitors 
of massive early-type galaxies that host the most massive SMBHs at low $z$.

Note that, although we have not included AGN feedback in the 
simulation, the galaxy merger remnant has an effective radius, central density, and stellar mass that agree well with observations of quiescent high-$z$ massive 
galaxies \citep{bezanson+09,Szomoru+12,bezanson+13}, including abundance matching constraints. Indeed, its star formation is efficiently quenched owing to the ``cosmological starvation'' mechanism described by \citet{feldmann+15}.

In order to have an idea how much off our timescales can be by not including AGN feedback, which could suppress star formation, we estimated the density of stars formed during the merger  within a sphere of 75 pc radius around the SMBH binary's center of mass. Most stars that interact with binary originate from distances within our chosen region (see figure \ref{fig3} right panel). 
Figure \ref{fig5} shows the total stellar surface density, as well as the contours in red of the surfaces density of the young stars with age below 22.5 Myr, when we switch between the hydro and the direct N-Body calculation. From the Figure it is evinced that
new stars form mostly in the central region.
We find that newly formed stars only contribute about 8\% to the total stellar density which would result in about 8\% smaller hardening rates \citep{seskh+15}. Hence even if AGN feedback shuts-off star formation completely, it would not cause any significant delay in the swift SMBH merger timescale found in this study.

However AGN feedback could still have a significant effect along earlier branches
of the galaxies merger trees on the simulated galaxies prior to the phase when the galaxy merger begins. While this is beyond the scope of our study, we can get some insight on this by comparing the effective radius of our galaxies with the effective radius of a much
larger sample of massive high redshift galaxies, which now includes star forming galaxies rather than only passive quiescent galaxies as in the \citet{bezanson+09,bezanson+13} work. Such a sample is that of the CANDELS survey \citep{papovich+15}, which extends to z = 2.8. When compared to the average effective radius of galaxies of the same stellar mass, the effective radius of our simulated galaxies is now about 50\% smaller.

This implies that our central stellar densities could be overestimated by a factor of 3, which would result in a SMBH merger timescale about a factor of 3 longer in the slowest phase of the decay, namely hardening due to 3-body encounters. 
This would imply a merger timescale close to 40 Myr, which would still be almost two
order of magnitude shorter than that reported by previous studies (e.g. \citet{khan+12b,vasiliev+15}).

We have presented only one multi-scale simulation due to the high computational cost that such calculations entail.
In order to understand how general is this result we revisited coalescence times obtained by a large suite of 
$N$-body simulations in \cite{khan+12b}.
We rescale\footnote{This rescaling is possible because the models by \citet{khan+12b} are scale-free and physical 
scaling can be obtained by comparing some characteristic length and mass of the model, e.g. influence radius and 
mass of SMBH to some reference values.} their merger product with the one obtained in our cosmological study. 
We find SMBH merger timescales from the time of binary formation in the range $10-30$~Myr, which satisfactorily 
match our results. 
This suggests that the timescale is primarily determined by the characteristic density of the host in the nucleus.
Since the stellar mass, $M_{\star}\lesssim10^{11}$~M$_{\odot}$, and the effective radius, $r_{\rm eff}\approx600$~pc, 
of our galaxy are typical of observed $z\gtrsim2$ massive early-type galaxies, 
we conclude that SMBH mergers in those systems at $z>2$ should be generically as fast as we find here.

In turn, the scaling argument suggests that the much longer coalescence timescales, of order a Gyr, 
should be the norm for massive, less dense early-type galaxies at low redshift. 
These are the galaxies hosting SMBHs with masses $>10^8$~M$_{\odot}$, whose
mergers should be detectable by Pulsar Timing Arrays (PTAs; \citealt{hobbs+10}). 
Recently, it has been argued that the lack of detection by PTAs might be difficult to reconcile 
with simple analytical predictions of the hardening rate which assume full loss cone and yield 
short SMBH merging timescales $\lesssim10^8$~yr \citep{shannon+15}. 
However, here we argue that such short coalescence timescales would occur only at high redshift, 
hence outside the observability window of PTAs.
At low redshift, coalescence times are of order of Gyr; the preceding galaxy merger phase up to SMBH 
binary formation will also be delayed
as major mergers become more rare, with less than 1 merger per galaxy per Gyr expected 
below $z=1$ for $L > L^{\star}$ galaxies \citep{stewart+09}.
Such a low merger rate would likely yield a GW background signal below the detection limit of PTAs, naturally 
explaining the current lack of detection.  

Nevertheless, early-type galaxies with properties expected for a recent dry merger (e.g. shells, tidal tails, and 
little recent SF) could be the ideal target for detecting SMBH binaries.
The fast coalescence times that we find support optimistic expectations for the number of GW emission events detectable with 
eLISA, at least for the most massive SMBHs in its detection window ($z 
~\sim 2-6$), in the range $10^6-10^8$~M$_{\odot}$.
The forecasts assume nearly instantaneous SMBH mergers after the galaxies merge \citep{amaro+13}.
Indeed, based on the known scaling relations, the host galaxies of such black holes should have stellar masses above 
$10^{9} - 10^{11}$~M$_{\odot}$, the mass range at which SF is most efficient in galaxies, yielding dense, triaxial merger 
remnants that would assist the prompt coalescence of their SMBHs.


\acknowledgments
We thank Robert Feldmann for providing us with the Argo simulation snapshots.
We thank Pedro R. Capelo and Alberto Sesana for useful discussions and for a thorough 
reading of the manuscript.
We thank Rainer Spurzem for support through the Silk Road Project at National Astronomical Observatories of Chinese Academy of Sciences.
F.M.K. acknowledges support by the Excellenzinitiative II ``Mobilit\"{a}tsma{\ss}nahmen im Rahmen internationaler Forschungskooperationen 2015-16'' of Heidelberg University.
D.F. is supported by the Swiss National Science Foundation under grant \#No. 200021\_140645.
P.B. is supported by
the Strategic Priority Research Program ``The Emergence of Cosmological Structure'' of the Chinese Academy of Sciences (No. XDB09000000), the Sonderforschungsbereich SFB 881 ``The Milky Way System'' (subproject Z2) of the German Research Foundation (DFG)and by the NASU under the Main Astronomical Observatory GRID/GPU computing cluster project.


\appendix

\section{Numerical tests of the particle splitting} \label{sec_appendix}

\begin{figure}
\epsscale{1.1}
\plotone{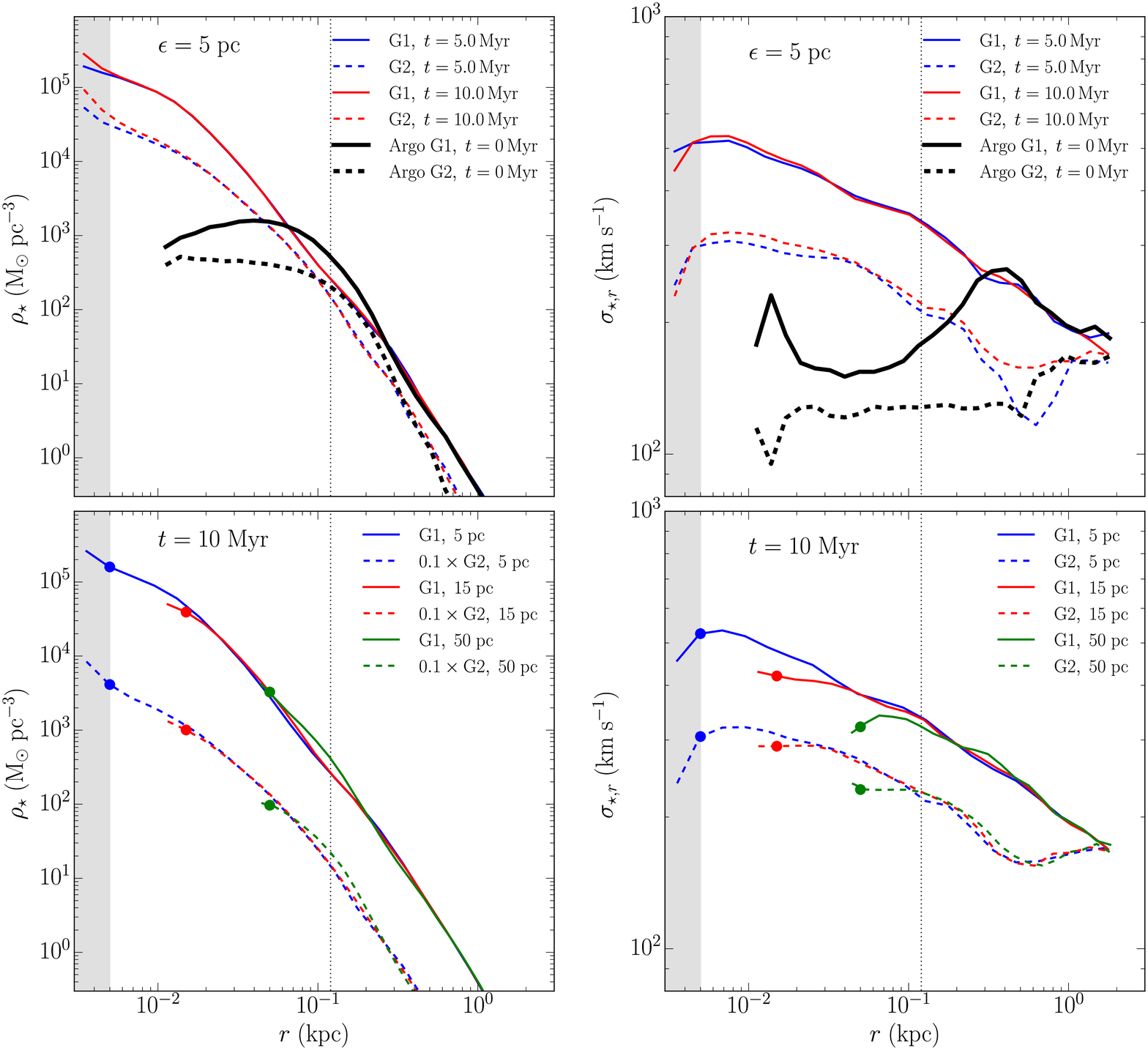}
\caption{Left and right columns show the stellar density and radial velocity dispersion profiles, respectively.
Upper row: evolution of the profiles of the central galaxy (G1, continuous lines) and the companion (G2, dashed lines) in the high resolution (5~pc) simulation at different times.
Thick lines (i.e. $t = 0$~Myr after particle splitting) shows the profile of the same galaxies in the original
Argo simulation.
Lower row: comparison of the profiles of the central galaxy (G1, continuous lines) and the companion (G2, dashed lines, decreased by a factor of 10 in $\rho_{\star}$ for clarity) after 10~Myr from simulations at different resolutions, namely 5 (red), 15 (green), and 50 (blue) pc.
The bullets indicate the softening of each run.
In all panels, the grey region marks 5~pc, while the vertical dotted line 120 pc.
\label{figA5}}
\end{figure}

\begin{figure}
\epsscale{0.65}
\plotone{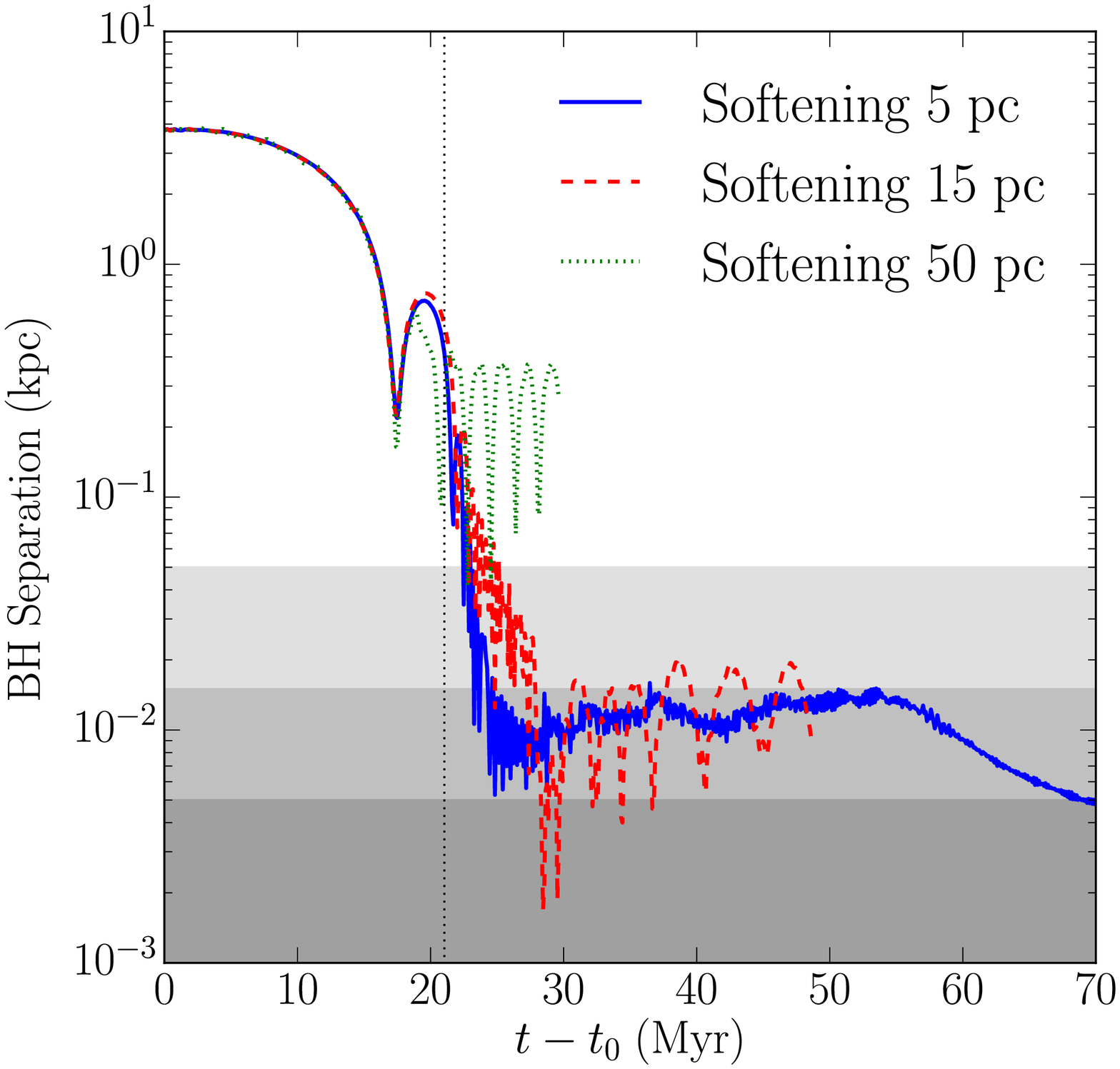}
\caption{Evolution with time of the separation between the two black holes at different spatial 
resolution in the re-sampled hydrodynamical simulations of the merger.
Blue solid, red dashed, and green dotted lines refer to $\epsilon=5$, 15 and 50~pc, respectively.
The same separations are also marked by the grey shaded bands, while the vertical line indicates the 
moment at which we initialize the direct $N$-body simulation.
\label{figA6}}
\end{figure}

We discuss convergence tests that we have performed in order to assess the numerical impact of the particle splitting.
We argue that this procedure perturbs the system only mildly and without altering the SMBH dynamics,
as already shown in previous work \citep{mayer+07,roskar+15}.
We run other two simulations of the galaxy merger phase at lower spatial resolution; 
we use a gravitational softening of 15 and 50~pc for stars, gas and SMBHs.
The upper row of Figure \ref{figA5} shows the combinend effect of particle splitting and of introducing 
the SMBHs at the galaxy centers.
Compared to the original cosmological simulation, the galaxies quickly re-adjust their density 
and velocity dispersion, steepening toward the center. 
This happens in a few Myr of evolution from the particle splitting; afterwards, they remain stable.
This is a natural effect since we add a new gravitational component and we increase the resolution.
Nonetheless, it is important to note that outside $\sim 2$ original softenings of Argo ($2 \epsilon = 240$~pc), 
the profiles do not change much, at least within $\sim 10$~Myr after the particle splitting, before than the 
galaxies start to strongly interact with each other.
This means that the particle splitting and the inclusions of the SMBHs do not introduce spurious effects on
scales larger than expected.
The lower row of Figure \ref{figA5} compares the density and velocity dispersion profiles at different 
numerical resolutions after the initial transient (i.e. 10~Myr after the particle splitting).
The profiles at different resolution mutually match each other outside $\sim 2$ times their own softening.
This clearly shows that the particle splitting is robust, at least down to the spatial resolution 
that we use in our production run.

As a final demonstration that the numerical procedure does not impact on our results, specifically on the 
SMBH dynamics, we compare in Figure \ref{figA6} the evolution of the SMBH separation at 
different resolutions. The SMBH separation is the same until the first pericenter; afterwards, 
the worse resolution case (50~pc) starts to deviate because the first pericenter occurs at a separation of 
$\sim 4$ softening lengths (gravity becomes Newtonian outside 2 softening lengths).
Instead, for better resolutions (i.e. smaller softening lengths), the SMBH 
dynamics converges until the separation reaches again $\sim 4$~softening lengths.
This is also because at higher resolution we can better resolve the formation of the stellar cusp around the 
SMBHs and their influence radii. In fact, in order to start a new, direct $N$-body simulation from the 
re-sampled merger, we use data at $t \sim 21.5$~Myr after the particle splitting, when simulations at 
different resolutions (at least 15 and 5 pc) converge and the different resolutions do not have an 
appreciable effect.


\end{document}